\documentclass[letter,twocolumn]{jpsj3}
\usepackage{txfonts}

\usepackage{amsmath,amssymb}

\usepackage[usenames]{color}
\usepackage{soul}

\usepackage[dvipdfmx]{graphicx}

\title{Ground-state and thermodynamic properties of an $S=1$ Kitaev model}

\author{Akihisa Koga, Hiroyuki Tomishige, and Joji Nasu}
\inst{Department of Physics, Tokyo Institute of Technology, Meguro, Tokyo 152-8551, Japan} 

\abst{
We study ground-state and thermodynamic properties of an $S=1$ Kitaev model.
We first clarify the existence of global parity symmetry in addition to
the local symmetry on each plaquette, which enables us
to perform the large scale calculations up to 24 sites.
It is found that the ground state should be singlet and
its energy is estimated as $E/N\sim -0.65J$,
where $J$ is the Kitaev exchange coupling.
We find that a lowest excited state belongs to
the same subspace as the ground state and the gap monotonically decreases
with increasing system size, which suggests that
the ground state of the $S=1$ Kitaev model is gapless.
By means of the thermal pure quantum states,
we clarify finite temperature properties characteristic of
the Kitaev models with $S\le 2$.
}

\begin{document}
\maketitle

Frustrated quantum spin systems have been 
the subjects of considerable interest.
One of the intriguing systems is the $S=1/2$ quantum spin Kitaev model
on the honeycomb lattice~\cite{Kitaev2006}, where
strong frustration emerges due to direction-dependent Ising interactions.
It is known that this model is exactly solvable and
its ground state is a quantum spin liquid state, where
spin degrees of freedom are decoupled by
itinerant Majorana fermions and
$Z_2$ fluxes~\cite{Kitaev2006,Feng2007,Chen2007,Chen2008}.
Two energy scales for distinct degrees of freedom
yield interesting finite temperature properties such as
double peak structure in the specific heat, spin dynamics,
and thermal Hall coefficients
at low temperatures~\cite{Nasu2015,Yamaji2016,Yoshitake2016,Nasu2017,suzuki2018pre,yamaji2018pre}.
Since magnetic properties of certain Mott insulators
with strong spin-orbit coupling should be described by the Kitaev model and its
extentions~\cite{Jackeli2009},
a lot of theoretical studies for these models~\cite{Chaloupka2013,Yamaji2014,Suzuki2015,PhysRevB.96.144414,Nakauchi,SuzukiYamaji} and
experimental studies
on candidates materials such as
$A_2$IrO$_3$($A$=Na, Li)~\cite{Singh2010,Singh2012,Comin2012,Choi2012},
$\alpha$-RuCl$_3$~\cite{Plumb2014,Kubota2015,Sears2015,Majumder2015}
and $\rm H_3LiIr_2O_6$~\cite{Bette2017}
have been done so far.

The key of these features characteristic of the Kitaev model
is the existence of the local conserved quantity
defined at each plaquette in the honeycomb lattice,
which is responsible for the absence of the long-range spin correlations.
From a fundamental viewpoint of quantum spin systems,
a question arises whether or not such interesting properties
appear in generalized Kitaev models with arbitrary spins.
It is known that
in the general spin $S$ Kitaev model~\cite{PhysRevB.78.115116},
there exist local $Z_2$ conserved quantities and
the ground state is nonmagnetic.
In spite of the presence of the local invariants, details of its excitations
and finite temperature properties remain unclear.
Very recently thermodynamic properties in the Kitaev model
with spin $S\le 5/2$
has been examined for an 8-site cluster~\cite{SuzukiYamaji}.
It has been found that,
in the specific heat, double peak structure appears
only for the $S=1/2$ and $S=1$ models, while a single peak appears
in the $S>1$ model.
However, the system size may not be large enough
to discuss thermodynamic properties for
the Kitaev models on the honeycomb lattice.
In addition, the role of the local conserved quantities was still unclear.
Therefore, it is necessary to clarify
how the $Z_2$ degrees of freedom inherent
in the Kitaev model affects
ground state and finite temperature properties.

In this Letter, we mainly focus on
the ground state and finite temperature properties of the $S=1$ Kitaev model.
First, we show the existence of the global parity symmetry
in addition to the $Z_2$ symmetry in each plaquette
in the system~\cite{PhysRevB.78.115116}.
Using these conserved quantities,
we examine ground state and finite-temperature properties in the large clusters up to 24 sites.
We clarify that ground state has no degeneracy and
belongs to the subspace with zero-fluxes, which is consistent with
the semiclassical results~\cite{PhysRevB.78.115116}.
We find that a lowest excited state belongs to
the same subspace as the ground state and the gap monotonically decreases
with increasing system size, which suggests that
the ground state of the $S=1$ Kitaev model is gapless.
Furthermore, using the thermal pure quantum (TPQ) state methods,
we demonstrate that the multiple peak structure appears
in the specific heat, similar to the $S=1/2$ Kitaev model~\cite{Nasu2015}.
Thermodynamic properties in the generic spin-$S$ Kitaev models with $S\le 2$
are also addressed.

\begin{figure}[htb]
\centering 
\includegraphics[width=6.5cm]{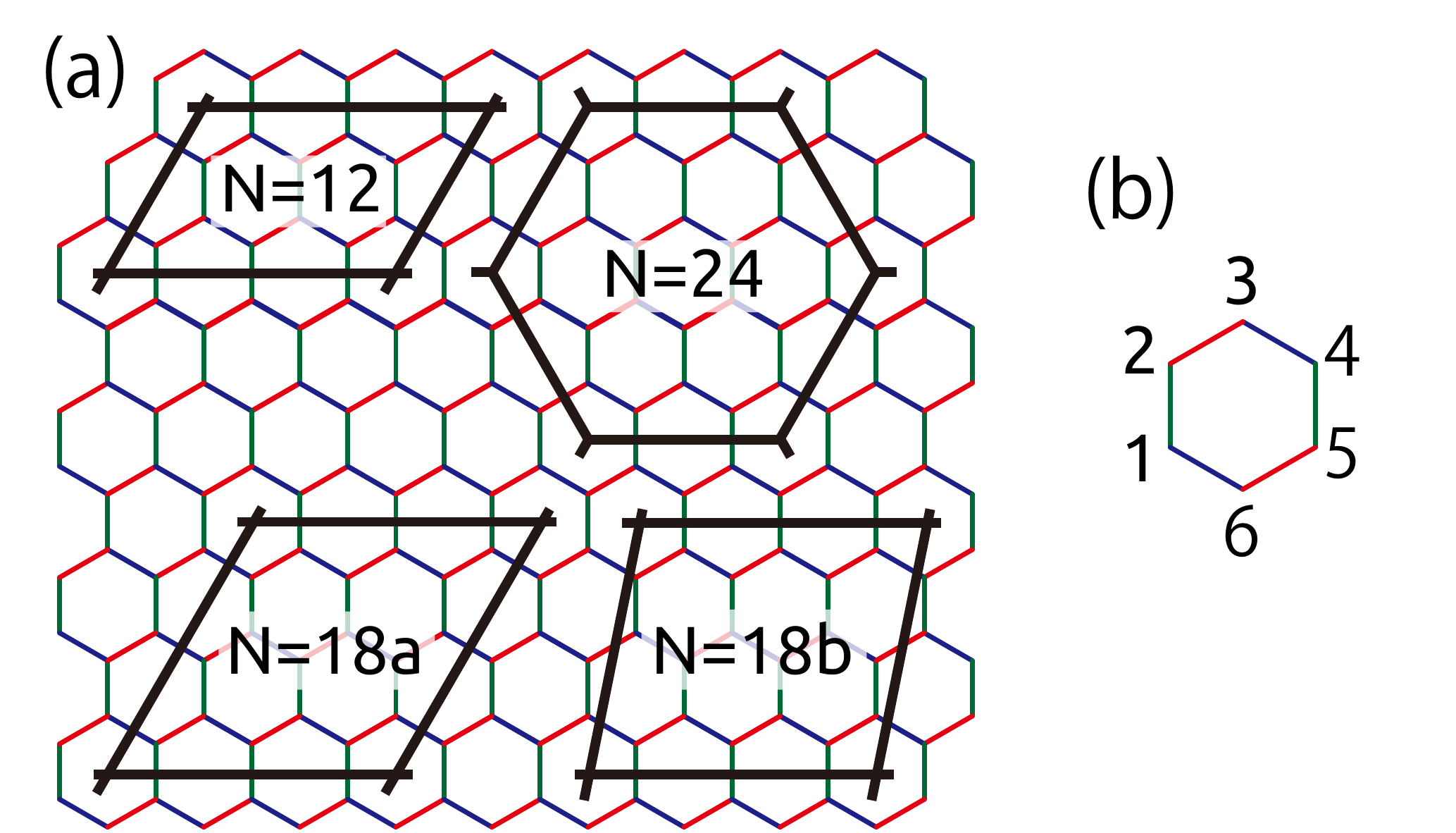}
\caption{
(a) Kitaev model on the honeycomb lattice.
Red, blue, and green lines denote $x, y$, and $z$ bonds, respectively.
Clusters used in the exact diagonalization and TPQ methods are represented
by the bold lines.
(b) Plaquette with sites marked $1-6$ is shown for
the corresponding operator $W_p$ defined in Eq.~(\ref{eq:Wp}).
}
\label{fig:model}
\end{figure}

We consider the $S=1$ Kitaev model given by
\begin{align}
H=-J\sum_{\langle i,j \rangle_x}S_i^xS_j^x
-J\sum_{\langle i,j \rangle_y}S_i^yS_j^y
-J\sum_{\langle i,j \rangle_z}S_i^zS_j^z,\label{eq:H}
\end{align}
where $S_i^\alpha$ is an $\alpha(=x,y,z)$ component of an $S=1$ spin operator
at the $i$th site in the honeycomb lattice.
The ferromagnetic interaction $J(>0)$ is defined
on three different types of the nearest
neighbor bonds, $x$ (red), $y$ (blue), and $z$ bonds (green),
respectively [see Fig. \ref{fig:model}(a)].
In this $S=1$ model, the basis sets $|\alpha\rangle\; (\alpha=x,y,z)$
are convenient~\cite{Tomishige}, which is explicitly given as
\begin{align}
|x\rangle &= -\frac{1}{\sqrt{2}}\left(|1\rangle -|-1\rangle \right),\\
|y\rangle &= \frac{i}{\sqrt{2}}\left(|1\rangle +|-1\rangle \right),\\
|z\rangle &= |0\rangle,
\end{align}
where $|m\rangle\; (m=-1, 0, 1)$ is an eigenstate of
the $z$-component of the spin operator $S^z$ with an eigenvalue $m$.
In this representation, we obtain the following relations as
\begin{align}
S^\alpha|\beta\rangle = i \epsilon_{\alpha\beta\gamma}|\gamma\rangle.
\label{eq:S}
\end{align}

Now, we consider the local Hamiltonian for the $x$ bond
between $i$ and $j$ sites, $H_{ij}^x=JS_i^xS_j^x$.
When a certain state $|\psi\rangle$ includes an $x$ state
in $i$th and/or $j$th sites,
$H_{ij}^x|\psi\rangle=0$ due to eq.~(\ref{eq:S}).
Then, nonzero components in the Hamiltonian $H_{ij}^x$ are explicitly given as
\begin{align}
H_{ij}^x|y_i z_j\rangle&=J|z_i y_j\rangle,\\
H_{ij}^x|z_i y_j\rangle&=J|y_i z_j\rangle,\\
H_{ij}^x|y_i y_j\rangle&=-J|z_i z_j\rangle,\\
H_{ij}^x|z_i z_j\rangle&=-J|y_i y_j\rangle.
\end{align}
We here note that the numbers of $y$ and $z$ states are not conserved,
but their parity is conserved.
When one considers all Kitaev interactions
in the system,
there exist global parity symmetries
for the number of $\alpha(=x,y,z)$ states.
In other word, the Kitaev Hamiltonian eq.~(\ref{eq:H}) commutes
the parity operators for $\alpha$ states as
\begin{align}
P_\alpha&=e^{i\pi n_\alpha}
=\exp\left[i\pi \sum_i \left(S_i^\alpha\right)^2\right],
\end{align}
where $n_\alpha(=\sum_i [1-(S_i^\alpha)^2 ])$ is the number operator
for $\alpha$ states, and
$P_x, P_y$, and $P_z$ commute with each other.
In addition, there exists a local conserved quantity
defined at each plaquette $p$ composed of the sites labeled as
$1, 2, \cdots, 6$ [see Fig.~\ref{fig:model}(b)]~\cite{PhysRevB.78.115116}:
\begin{align}
W_p&=\exp\Big[i\pi \left(S_1^x+S_2^y+S_3^z+S_4^x+S_5^y+S_6^z\right)\Big].
\label{eq:Wp}
\end{align}
This operator commutes with not only $W_{p'}$ on another plaquette $p'$ and
the Hamiltonian eq.~(\ref{eq:H}), but also the parity operators $P_\alpha$.
Then, the Hilbert space of the Hamiltonian can be classified into
each subspace specified by ${\cal S}=[\{p_\alpha\}; \{w_p\}]$,
where $p_\alpha(=\pm 1)$ and $w_p(=\pm 1)$ are eigenvalues of the operators
$P_\alpha$ and $W_p$.
This enables us to perform the numerical calculations for large clusters
in the Lanczos and TPQ state methods.
For example, the Hilbert space in the cluster with $N=24$ is divided into
2,048 subspaces, in which the maximum rank of 
the block-diagonalized Hamiltonian is 140,279,025.
Applying the Lanczos and TPQ state methods to the systems
with periodic boundary conditions, we discuss ground state and
finite temperature properties.



\begin{figure}[htb]
\includegraphics[width=9cm]{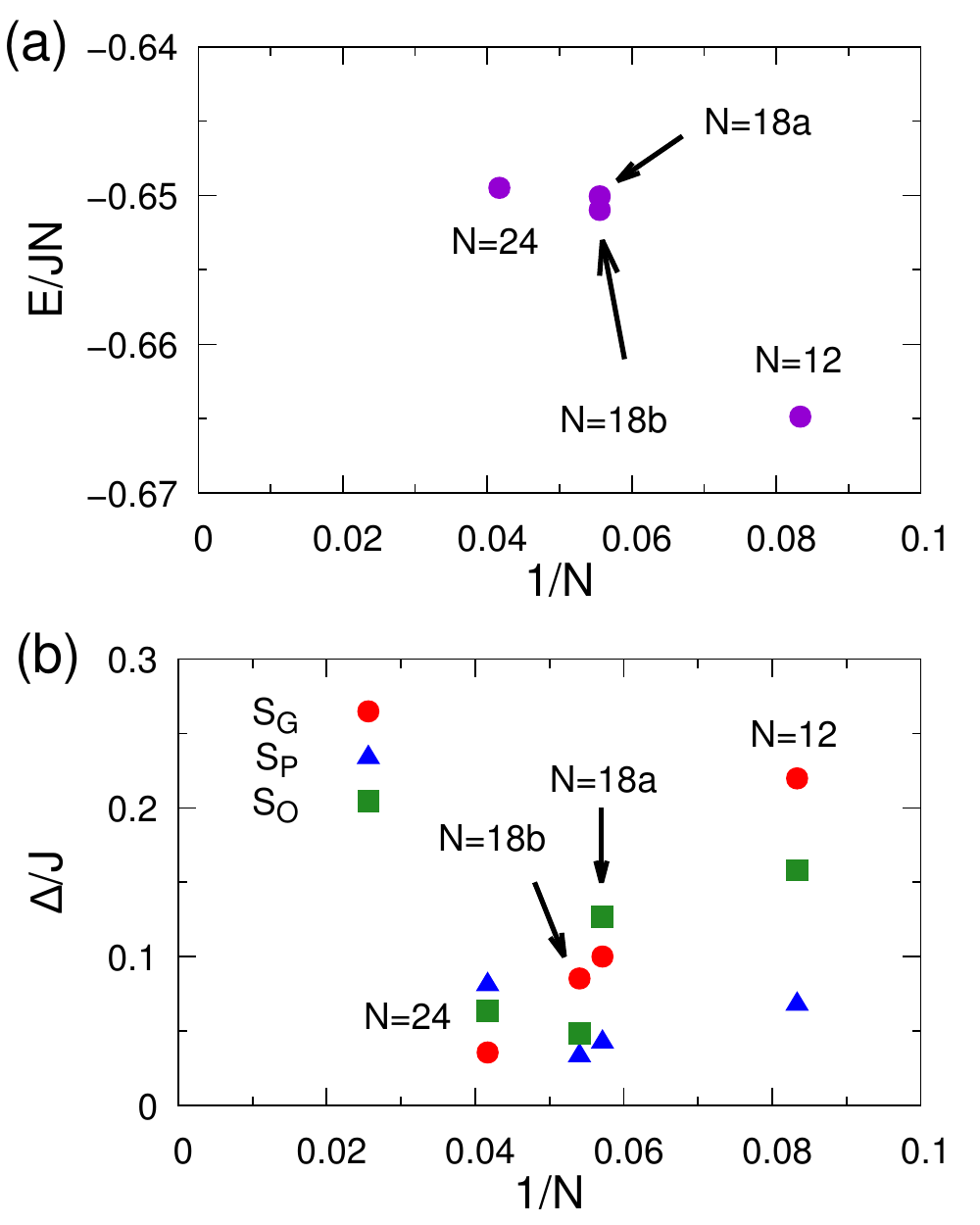}
\caption{
(a) Ground state energy as a function of $1/N$.
(b) Excitation gap as a function of $1/N$.
Circles, triangles, and squares represent the gap between the ground state
and lowest excited state in the subspaces ${\cal S}_G, {\cal S}_P$, and ${\cal S}_O$,
respectively.
For convenience, we have plotted the results for $N=18a$ and $N=18b$
with tiny offsets.
}
\label{fig:ene}
\end{figure}

In our study, we deal with the $S=1$ Kitaev model
on the clusters with $N=12, 18$, and 24,
which are shown in Fig.~\ref{fig:model}(a).
Calculating the lowest energy for each cluster,
we find that the ground state has no degeneracy and
always belongs to the subspace ${\cal S}_G=[\{p_\alpha=+1\}, \{w_p=+1\}]$,
corresponding to the zero-flux sector.
This is consistent with the results obtained from
the semiclassical approach~\cite{PhysRevB.78.115116}.
The ground state energy for each cluster is shown in Fig.~\ref{fig:ene}(a).
In the thermodynamic limit, the ground state energy appears to be given by
$E/N\sim -0.65J$
although the systematic finite size scaling
is necessary to be deduced quantitatively.

We also discuss the elementary excitations in the $S=1$ Kitaev model,
examining all subspaces for the clusters.
Figure~\ref{fig:ene}(b) shows lowest excitation energies in the subspaces
${\cal S}_G, {\cal S}_P=[\{p_\alpha=+1\}]-{\cal S}_G$
and ${\cal S}_O=[\{p_\alpha=+1\}]^c$,
where $[\{p_\alpha\}]$ is the subspace with eigenvalues ${p_\alpha}$,
and ${\cal S}^c$ means the complement space of ${\cal S}$.
The excited states strongly depend on the size and/or shape of clusters,
as shown in Fig.~\ref{fig:ene}(b).
Therefore, we believe that the largest cluster $N=24$ with isotropic geometry
should capture the nature of the excitations in the thermodynamic limit.
In the $N=24$ system, the lowest excited state
belongs to the subspace ${\cal S}_G$,
which is the same subspace for the ground state.
Furthermore, we find that
the excitation gap in the subspace ${\cal S}_G$
monotonically decreases with increasing the system size
and is much smaller than the exchange constant $J$.
These suggest that in the thermodynamic limit,
the $S=1$ Kitaev model has gapless excitations in the same subspace
${\cal S}_G$.
This is similar to the case with the $S=1/2$ Kitaev model,
where the gapless dispersion for Majorana fermions is obtained
in the subspace corresponding to the zero-flux sector~\cite{Kitaev2006}.

\begin{figure}[htb]
\includegraphics[width=9cm]{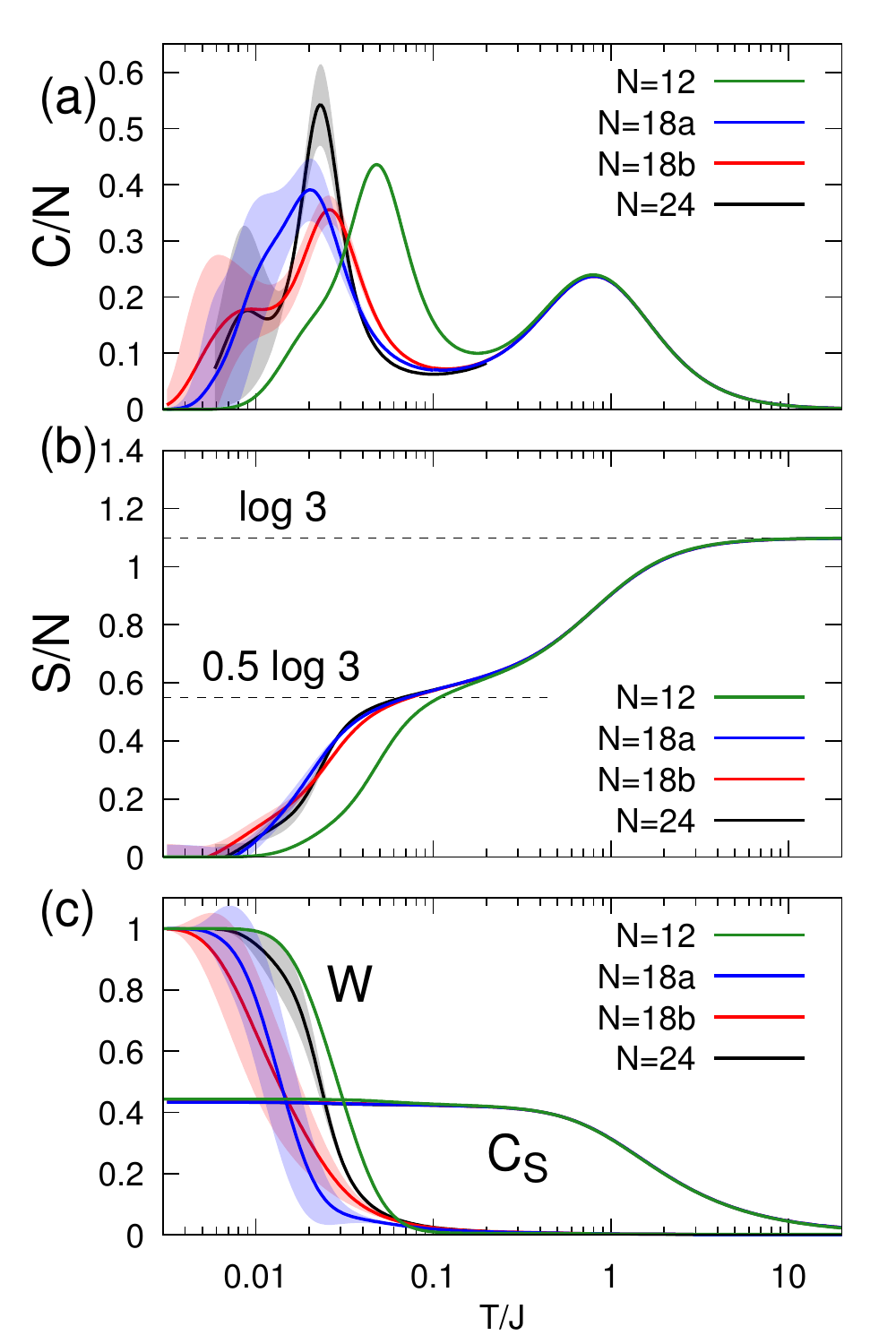}
\caption{
(a) Specific heat, (b) entropy, (c) the expectation values of
local conserved quantity and nearest-neighbor spin correlations
as a function of the temperature for the systems.
Shaded areas 
are estimated
by using standard deviation of the results obtained from TPQ states.
}
\label{fig:tpq}
\end{figure}

Next, we examine thermodynamic properties in the $S=1$ Kitaev model.
We calculate thermodynamic quantities for a small cluster with $N=12$,
obtaining all eigenenergies by means of the exact diagonalization method.
Furthermore, we make use of the TPQ states for larger clusters.
As for two kinds of clusters with $N=18$,
we evaluate thermodynamic quantities and expectation values
by means of hundred independent TPQ states.
By contrast, the numerical cost is expensive for the $N=24$ site cluster,
and only two TPQ states are treated.
The obtained results are shown in Fig.~\ref{fig:tpq}.
We clearly find in Fig.~\ref{fig:tpq}(a),
the multiple peak structure in the specific heat $C$,
which is consistent with the previous work~\cite{SuzukiYamaji}.
Now, we focus on the peak structure around $T=T_H\sim 0.8J$.
Since the specific heat around $T_H$ is almost independent of the cluster size,
this peak structure appears in the thermodynamic limit.
On the other hand, low temperature structure is sensitive to the cluster size,
which originates from the fact that
low-energy excitations depend on the size and/or shape of clusters
[see Fig.~\ref{fig:ene}(b)].
Nevertheless, the peak appears around $T_L/J\sim 0.02$
except for the smallest cluster with $N=12$,
which suggests that the low temperature peak indeed appears
in the thermodynamic limit.
In addition to these features, the shoulder or peak structure appears
in lower energy region
$T/J\sim 0.009$, as shown in Fig.~\ref{fig:tpq}(a).
Since it strongly depends on the clusters,
further detailed analysis with larger clusters is necessary
to clarify whether or not such low-temperature behavior appears in the thermodynamic limit.

We also calculate the entropy $S$, and
the expectation values of the local conserved quantity
$W$, and spin-spin correlations $C_s$, which are defined as
\begin{align}
W&=\frac{1}{N_p}\sum_p \langle W_p \rangle,\\
C_s&=\frac{1}{N_B}\sum_{\langle i,j\rangle_\alpha} \langle S_i^\alpha S_j^\alpha\rangle, 
\end{align}
where $N_p(=N/2)$ is the number of plaquettes, and
$N_B(=3N/2)$ is the number of bonds.
Note that $C_s$ is proportional to the energy as $C_s=-E/(JN_B)$.
The results are shown in Figs.~\ref{fig:tpq}(b) and (c).
It is found that,
decreasing temperatures,
nearest-neighbor spin-spin correlations develop around $T=T_H$ and
the plateau behavior appears in the curve of the entropy 
at $S/N\sim 1/2\log 3$ around $T/J\sim 0.1$.
On the other hand, the expectation value $W$ changes rapidly
at lower temperature $T_L/J\sim 0.02$.
Therefore, the peaks in the specific heat at $T=T_H$ and $T_L$
should be related to spin-spin correlations and $Z_2$ local conserved quantity.

From the above discussions, we have confirmed that, in the $S=1$ Kitaev model,
at least, double peak structure appears in the specific heat.
This originates from two distinct temperature dependences
for nearest-neighbor spin-spin correlations and the local conserved quantity,
which are essentially the same as those
in the $S=1/2$ Kitaev model~\cite{Nasu2015}.
Then, similar finite-temperature behavior is naively expected
even in larger spin $S$ models
since they have the local conserved quantity~\cite{PhysRevB.78.115116}.
To clarify this, we also examine finite-temperature properties for
the Kitaev models with $S=3/2$ and $S=2$,
using the TPQ states for the clusters with $N=16$ and $N=12$, respectively.
First, we have compared the results for the $N=12$ and $N=16$ clusters
in the spin $S=3/2$ system, and have confirmed that
the higher temperature peak is not changed in the specific heat.
Therefore, we believe that 
the $N=12$ site calculations can capture higher temperature peak 
in the thermodynamic limit.
Meanwhile, the characteristic temperature for low temperature peak
in the specific heat
depends on the system size,
which is similar to the $S=1$ case (see Fig.~\ref{fig:tpq}).

\begin{figure}[htb]
\includegraphics[width=8cm]{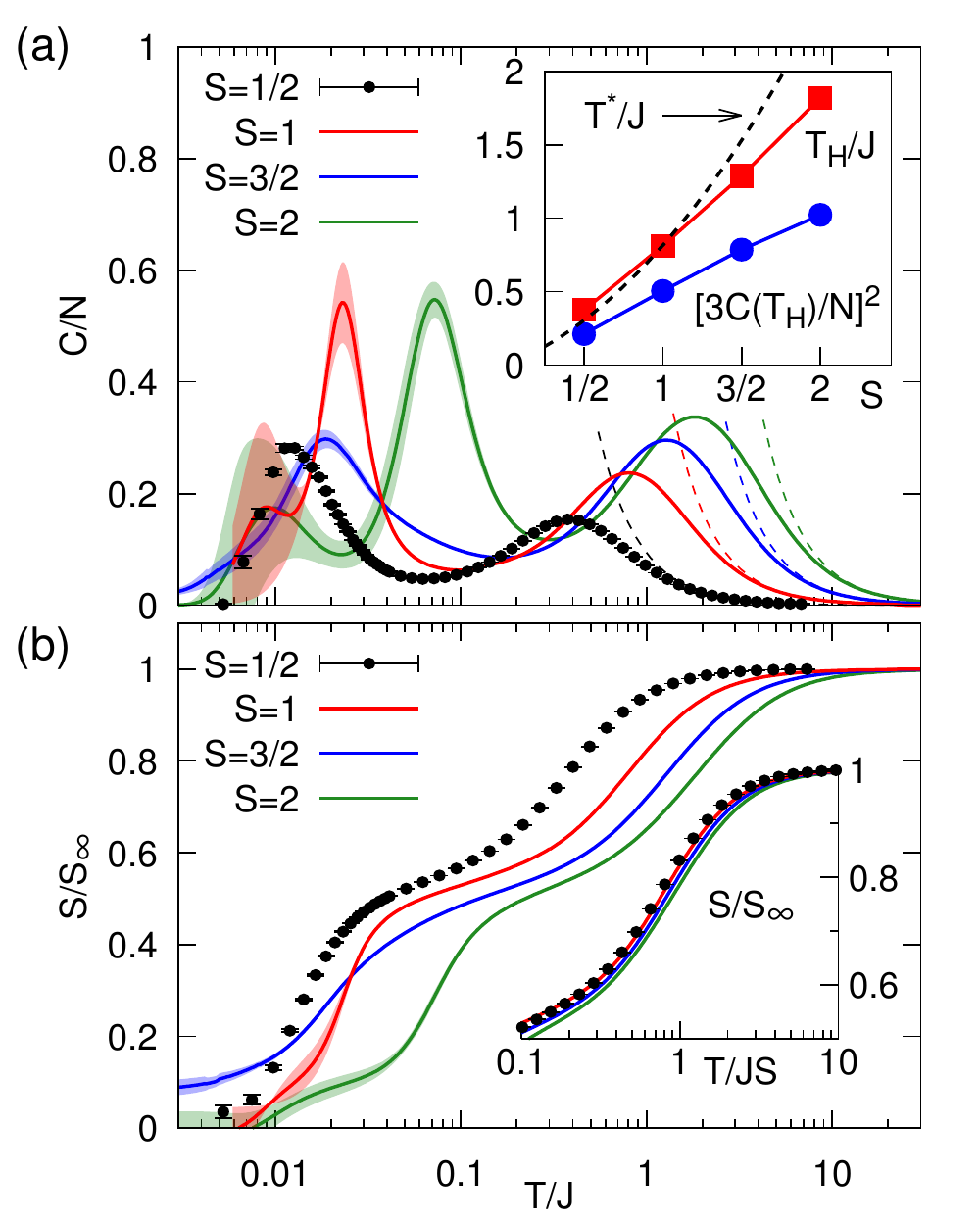}
\caption{
(a) Specific heat and (b) normalized entropy
as a function of the temperature for the Kitaev model with spin $S\le 2$,
where $S_\infty=N\log (2S+1)$.
Shaded areas are estimated by using standard deviation of the results
obtained from initial random states.
Data with error bars for $S=1/2$ Kitaev model has been obtained
by the Monte Carlo simulations with $N=800$ sites~\cite{Nasu2015,Nasu2017}.
Dashed lines represent the high-temperature behavior $C= (T^*/T)^2$,
where  $T^*=JS(S+1)/\sqrt{6}$.
Inset of the figure (a) shows the spin dependence of
the characteristic temperatures $T_H$ and $T^*$
and the square of its specific heat $C$.
Inset of the figure (b) shows the entropy as a function of $T/(JS)$.
}
\label{fig:tpqS}
\end{figure}
Figure~\ref{fig:tpqS} shows the specific heat and entropy
in the Kitaev models with $S\le 2$.
Namely, the $S=1/2$ results are obtained by the Monte Carlo method
for $N=800$ sites~\cite{Nasu2015,Nasu2017}.
The results for $S=1, 3/2$ and $2$ are obtained by the TPQ state methods
for $N=24$, $16$ and $12$ clusters, respectively.
We find multiple peak structures in the specific heat
for the spin-$S$ Kitaev model in Fig.~\ref{fig:tpqS}(a).
This is contrast to the previous work~\cite{SuzukiYamaji},
where a single peak is observed in the specific heat
for the $S>1$ cases in an 8-site cluster.
This should originate from the fact that
the system size treated in previous works is too small
to discuss finite temperature properties in the thermodynamic limit.
Now, we focus on the higher-temperature peak in the specific heat,
which can be described quantitatively in our methods.
It is found that the characteristic temperature and the height of the peak
monotonically increase with increasing the amplitude of spin $S$,
as shown in Fig.~\ref{fig:tpqS}(a).
To clarify the spin dependence,
we also plot its characteristic temperature $T_H$ and
the value of the specific heat at $T_H$
in the inset of Fig.~\ref{fig:tpqS}(a).
We numerically find that these quantities are almost scaled as
$T_H/J \propto S$ and $C(T_H)/N\propto \sqrt{S}$.
These are in contrast to the spin dependence of
the characteristic temperature $T^*$ derived from
the high-temperature expansion:
the specific heat is expanded as $C/N = (T^*/T)^2 + O(T^{-3})$
with $T^*=JS(S+1)/\sqrt{6}$.
The difference in the characteristic temperatures
implies that the fractionalization of spin occurs
in the spin-$S$ Kitaev models 
as well as the $S=1/2$ Kitaev model~\cite{Kitaev2006}.
To confirm this, we calculate the temperature dependence of
the entropy, as shown in Fig.~\ref{fig:tpqS}(b).
We find that the plateau appears in the curve of the entropy at
$S/S_\infty\sim 1/2$,
suggesting that spin with the amplitude $S$ is fractionalized
into two halves with distinct energy scales.
This is corroborated by the fact that the temperature of normalized entropy
$S/S_\infty$ is well scaled by $JS$,
as shown in the inset of Fig.~\ref{fig:tpqS}(b).
We naively expect that the decrease of temperatures induces 
another peak in the specific heat, where the residual entropy is released.
Unfortunately, the low-temperature peak could not be discussed quantitatively
since the clusters treated in the present study are not enough large
to describe low energy excitations correctly.
Therefore, the systematic analysis with larger clusters are needed
to clarify whether peak and/or shoulder structures appears or not
in the lower temperature region,
which is left for a future work.


In summary, we have studied an $S=1$ Kitaev model
to discuss ground-state and finite temperature properties.
We have demonstrated the existence of parity symmetry as well as
the local conserved quantity, which enables us
to perform the calculations in large clusters up to twenty-four sites.
We have found that the ground state is singlet and
its energy is estimated as $E/N\sim -0.65$.
In addition, a lowest excited state belongs to
the same subspace as the ground state and the gap monotonically decreases
with increasing system size, which suggests that
the ground state of the $S=1$ Kitaev model is gapless.
By means of the TPQ states,
we have examined finite temperature properties and clarified that
the multiple peak structure in the specific heat appears in the Kitaev models
with spin $S\le 2$.
This should be a common feature of the general $S$ Kitaev model,
which might originate from the presence of a local conserved quantity
on each plaquette in the lattice.

This work is supported by Grant-in-Aid for Scientific Research from
JSPS, KAKENHI Grant Number JP17K05536, JP16H01066 (A.K.) and
JP16K17747, JP16H02206, JP16H00987 (J.N.).
Parts of the numerical calculations were performed
in the supercomputing systems in ISSP, the University of Tokyo.

\bibliographystyle{jpsj}
\bibliography{./refs}

\begin{thebibliography}{10}

\bibitem{Kitaev2006}
A.~Kitaev: Ann. Phys. (N. Y.) {\bfseries 321} (2006) 2.

\bibitem{Feng2007}
X.-Y. Feng, G.-M. Zhang, and T.~Xiang: Phys. Rev. Lett. {\bfseries 98} (2007)
  087204.

\bibitem{Chen2007}
H.-D. Chen and J.~Hu: Phys. Rev. B {\bfseries 76} (2007) 193101.

\bibitem{Chen2008}
H.-D. Chen and Z.~Nussinov: J. Phys. A {\bfseries 41} (2008) 075001.

\bibitem{Nasu2015}
J.~Nasu, M.~Udagawa, and Y.~Motome: Phys. Rev. B {\bfseries 92} (2015) 115122.

\bibitem{Yamaji2016}
Y.~Yamaji, T.~Suzuki, T.~Yamada, S.-i. Suga, N.~Kawashima, and M.~Imada: Phys.
  Rev. B {\bfseries 93} (2016) 174425.

\bibitem{Yoshitake2016}
J.~Yoshitake, J.~Nasu, and Y.~Motome: Phys. Rev. Lett. {\bfseries 117} (2016)
  157203.

\bibitem{Nasu2017}
J.~Nasu, J.~Yoshitake, and Y.~Motome: Phys. Rev. Lett. {\bfseries 119} (2017)
  127204.

\bibitem{suzuki2018pre}
T.~Suzuki and S.-i. Suga:   (unpublished) arXiv:1802.00545.

\bibitem{yamaji2018pre}
Y.~Yamaji, T.~Suzuki, and M.~Kawamura:   (unpublished) arXiv:1802.02854.

\bibitem{Jackeli2009}
G.~Jackeli and G.~Khaliullin: Phys. Rev. Lett. {\bfseries 102} (2009) 017205.

\bibitem{Chaloupka2013}
J.~c.~v. Chaloupka, G.~Jackeli, and G.~Khaliullin: Phys. Rev. Lett. {\bfseries
  110} (2013) 097204.

\bibitem{Yamaji2014}
Y.~Yamaji, Y.~Nomura, M.~Kurita, R.~Arita, and M.~Imada: Phys. Rev. Lett.
  {\bfseries 113} (2014) 107201.

\bibitem{Suzuki2015}
T.~Suzuki, T.~Yamada, Y.~Yamaji, and S.-i. Suga: Phys. Rev. B {\bfseries 92}
  (2015) 184411.

\bibitem{PhysRevB.96.144414}
R.~R.~P. Singh and J.~Oitmaa: Phys. Rev. B {\bfseries 96} (2017) 144414.

\bibitem{Nakauchi}
A.~Koga, S.~Nakauchi, and J.~Nasu: preprint  (unpublished) arXiv:1705.09659.

\bibitem{SuzukiYamaji}
T.~Suzuki and Y.~Yamaji: Physica B  (in press.).

\bibitem{Singh2010}
Y.~Singh and P.~Gegenwart: Phys. Rev. B {\bfseries 82} (2010) 064412.

\bibitem{Singh2012}
Y.~Singh, S.~Manni, J.~Reuther, T.~Berlijn, R.~Thomale, W.~Ku, S.~Trebst, and
  P.~Gegenwart: Phys. Rev. Lett. {\bfseries 108} (2012) 127203.

\bibitem{Comin2012}
R.~Comin, G.~Levy, B.~Ludbrook, Z.-H. Zhu, C.~N. Veenstra, J.~A. Rosen,
  Y.~Singh, P.~Gegenwart, D.~Stricker, J.~N. Hancock, D.~van~der Marel, I.~S.
  Elfimov, and A.~Damascelli: Phys. Rev. Lett. {\bfseries 109} (2012) 266406.

\bibitem{Choi2012}
S.~K. Choi, R.~Coldea, A.~N. Kolmogorov, T.~Lancaster, I.~I. Mazin, S.~J.
  Blundell, P.~G. Radaelli, Y.~Singh, P.~Gegenwart, K.~R. Choi, S.-W. Cheong,
  P.~J. Baker, C.~Stock, and J.~Taylor: Phys. Rev. Lett. {\bfseries 108} (2012)
  127204.

\bibitem{Plumb2014}
K.~W. Plumb, J.~P. Clancy, L.~J. Sandilands, V.~V. Shankar, Y.~F. Hu, K.~S.
  Burch, H.-Y. Kee, and Y.-J. Kim: Phys. Rev. B {\bfseries 90} (2014) 041112.

\bibitem{Kubota2015}
Y.~Kubota, H.~Tanaka, T.~Ono, Y.~Narumi, and K.~Kindo: Phys. Rev. B {\bfseries
  91} (2015) 094422.

\bibitem{Sears2015}
J.~A. Sears, M.~Songvilay, K.~W. Plumb, J.~P. Clancy, Y.~Qiu, Y.~Zhao,
  D.~Parshall, and Y.-J. Kim: Phys. Rev. B {\bfseries 91} (2015) 144420.

\bibitem{Majumder2015}
M.~Majumder, M.~Schmidt, H.~Rosner, A.~A. Tsirlin, H.~Yasuoka, and M.~Baenitz:
  Phys. Rev. B {\bfseries 91} (2015) 180401.

\bibitem{Bette2017}
S.~Bette, T.~Takayama, K.~Kitagawa, R.~Takano, H.~Takagi, and R.~E. Dinnebier:
  Dalton Trans. {\bfseries 46} (2017) 15216.

\bibitem{PhysRevB.78.115116}
G.~Baskaran, D.~Sen, and R.~Shankar: Phys. Rev. B {\bfseries 78} (2008) 115116.

\bibitem{Tomishige}
H.~Tomishige, J.~Nasu, and A.~Koga: Phys. Rev. B {\bfseries 97} (2018) 094403.

\end{thebibliography}

\end{document}